\documentclass[aps,pre,amsmath,amsfonts,twocolumn,longbibliography,superscriptaddress,floatfix]{revtex4-1}
\usepackage{graphicx}
\usepackage{dcolumn}
\usepackage{bm}
\usepackage{url}
\usepackage{amsmath}
\usepackage{amssymb}

\usepackage{natbib}
\usepackage{nicefrac}
\usepackage{xcolor, soul} 

\begin{document}


\title{Keeping it Together: Interleaved \textbf{\textsl{Kirigami}} Extension Assembly} 

\author{Xinyu Wang}
\affiliation{School of Civil Engineering, Southeast University, Nanjing, 210096, China}
\affiliation{Department of Physics and Astronomy, University of Pennsylvania, Philadelphia PA 19104, USA}
\author{Simon D. Guest}
\affiliation{Department of Engineering, University of Cambridge, Trumpington Street, Cambridge CB2 1PZ, UK}
\author{Randall D. Kamien}
\affiliation{Department of Physics and Astronomy, University of Pennsylvania, Philadelphia PA 19104, USA}
\begin{abstract}
Traditional {\sl origami} structures can be continuously deformed back to a flat sheet of paper, while traditional {\sl kirigami} requires glue or seams in order to maintain its rigidity.  In the former, non-trivial geometry can be created through overfolding paper while, in the latter, the paper topology is modified.  Here we propose a hybrid approach that relies upon overlapped flaps that create in-plane compression resulting in the formation of polyhedra composed of freely-supported plates.  Not only are these structures self-locking, but they have colossal load-to-weight ratios of order $10^4$. 
\end{abstract}

\maketitle
\section{Introduction}
The role of self- folding and unfolding has become ever more used as a framework to understand natural structure~\cite{ml,sabetta}.  Since the tunability of geometry is scale-invariant, {\sl origami} and {\sl kirigami} inspired structures translate from large to small scales as can be seen from the packaging materials in daily life~\cite{wrap} to deployable solar panels for space missions~\cite{miura,solar}.  Moreover, their transformability serves as a powerful tool to program shape-induced properties: these architected structures have been utilized in flexible electronics~\cite{john}, mechanical metamaterials~\cite{tunastiff,pau,kww}, and soft robots~\cite{berto,cohen}. 
  
In practice, {\sl origami} and {\sl kirigami}  start with geometric design.  However, paper, time, and human effort are all required in order to design {\sl origami} structures with exquisitely detailed Gaussian curvature \cite{tachi}.  Recently,  {\sl kirigami} methods have been developed that allow, in addition to folding, cutting and rejoining that drastically reduce the complexity of the inverse design problem~\cite{paper1,paper2,paper3}.  By maintaining a lattice structure and inducing Gaussian curvature via buckling of (two-dimensional) disclinations into the third dimension, these lattice {\sl kirigami} methods provide an algorithmic approach to design that maintains edge lengths on the lattice and dual lattice. Although the {\sl kirigami} rules provide much simpler ways to introduce curvature, we still need glue, seams, or a zipper to rejoin the shape (similarly in the of case {\sl origami} if we want to hold the shape at specific folding angles).  In comparison to the existing library of  \textsl{origami} and \textsl{kirigami} motifs, here we develop more possibilities in terms of self-locking, high stiffness structures, improving the ability to tune  material strength and anisotropy.

\begin{figure}[t]
 	\includegraphics[width=0.9\linewidth]{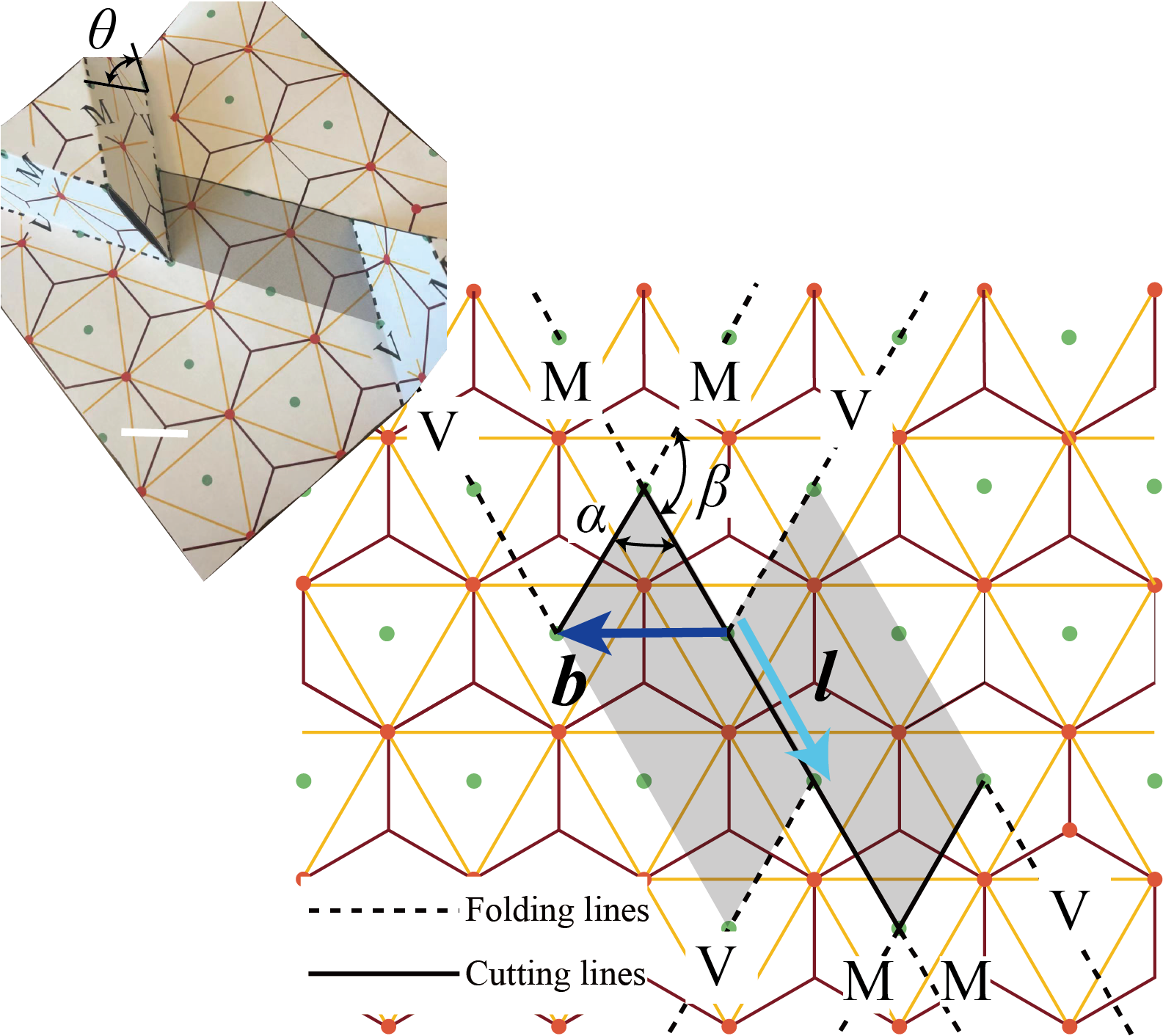}
 	\caption{\label{fig:basicmotif}
 		A diagram of our basic motif.  We cut along solid lines and make mountain (M) and valley (V) folds along the dashed lines. In the final state, the hatched areas overlap. In the inset we show the assembled structure. (scale bar: 1cm)}
 \end{figure}

Locking mechanisms have been considered by stacking identical models to enhance rigidity~\cite{kww,guest}.  In this paper we demonstrate a novel way of self-locking that is a hybrid of {\sl origami} and {\sl kirigami}. Specifically, instead of excising material from a dislocation~\cite{paper1,paper2}, we make cuts without removing any material. 
Upon folding, a dislocation-antidislocation pair includes a pair of interleaved flaps. Different from complex \textsl{origami} with overlapped parts underneath, we use the overlapped areas serving as intrinsic locks rather than just hiding and wasting the material(Fig.\ \ref{fig:basicmotif}). 
 
\textsl{Origami} and \textsl{kirigami} designs ({\sl e.g.}, eggcrate and honeycomb patterns) have been used effectively for building robust structures ~\cite{pau2, youzhong, mark}. However, most of those structures, for example eggcrate \textsl{origami}, are not flat-foldable or reversible -- the summation of their vertex angles is less than $2\pi$. To fabricate those structures, we must either use complicated folds or first cut those parts and  glue them together in order to hold up the whole structure. The latter can largely weaken the stiffness of eggcrate \textsl{origami}. Even Miura-ori inspired designs, that are supposed to be both rigid-foldable and flat-foldable, still need external locks or synchrononous folding processes to bear weight~\cite{mark}. However, our proposed assembly provides a novel locking method that can transfer the structure between a bearing structure and a piece of flat paper. The assembled triangular lattice exhibits a surprisingly strong stiffness.

Our pattern has twofold, threefold, and sixfold symmetry. For anisotropic materials (like paper), the  high symmetry can neutralize the anisotropy of the material and create  a direction-invariant structure. A six-unit loop (see Fig.\ \ref{fig:unitassemb}) makes a compromise between six different bending directions, rendering an isotropic bending modulus to quadratic order. For potential applications, our design lends itself to deployable structures like portable shelters, architectural canopies, and furniture. In analogy with {\sl origami}-inspired energy absorption structures~\cite{youzhong}, it should also be possible to stack the proposed pattern into multiple layers and harness the buckling behaviors of sidewalls to absorb energy.

In this paper, we first demonstrate the extended assembly method {for a special locking mechanism}. We modify the topology of the flat-state lattice and arrange dislocation pairs to form a new pattern. We then explore its working mechanism by strength tests. By observing pre-buckling and post-buckling behaviors of different models, we demonstrate the influence of ``flaps'' and neighbors.  We find that our motif in honeycomb lattice (or triangular lattice) is unusually strong and hypothesize that the geometry and mechanics conspire to create stable, ``virtual'' polyhedra, held together by in-plane compression and {\sl helped by }friction in the initial stage of deformation.  We test this hypothesis in numerous way by studying related cutting motifs, modifications to our original design, and { numerical finite-element models (FEM). Finally, we apply this motif to different materials with differing roughness and establish the role of friction in maintaining the unstressed structure.}   

\section{Extended Assembly}

\begin{figure}[h]
	\includegraphics[width=1\linewidth]{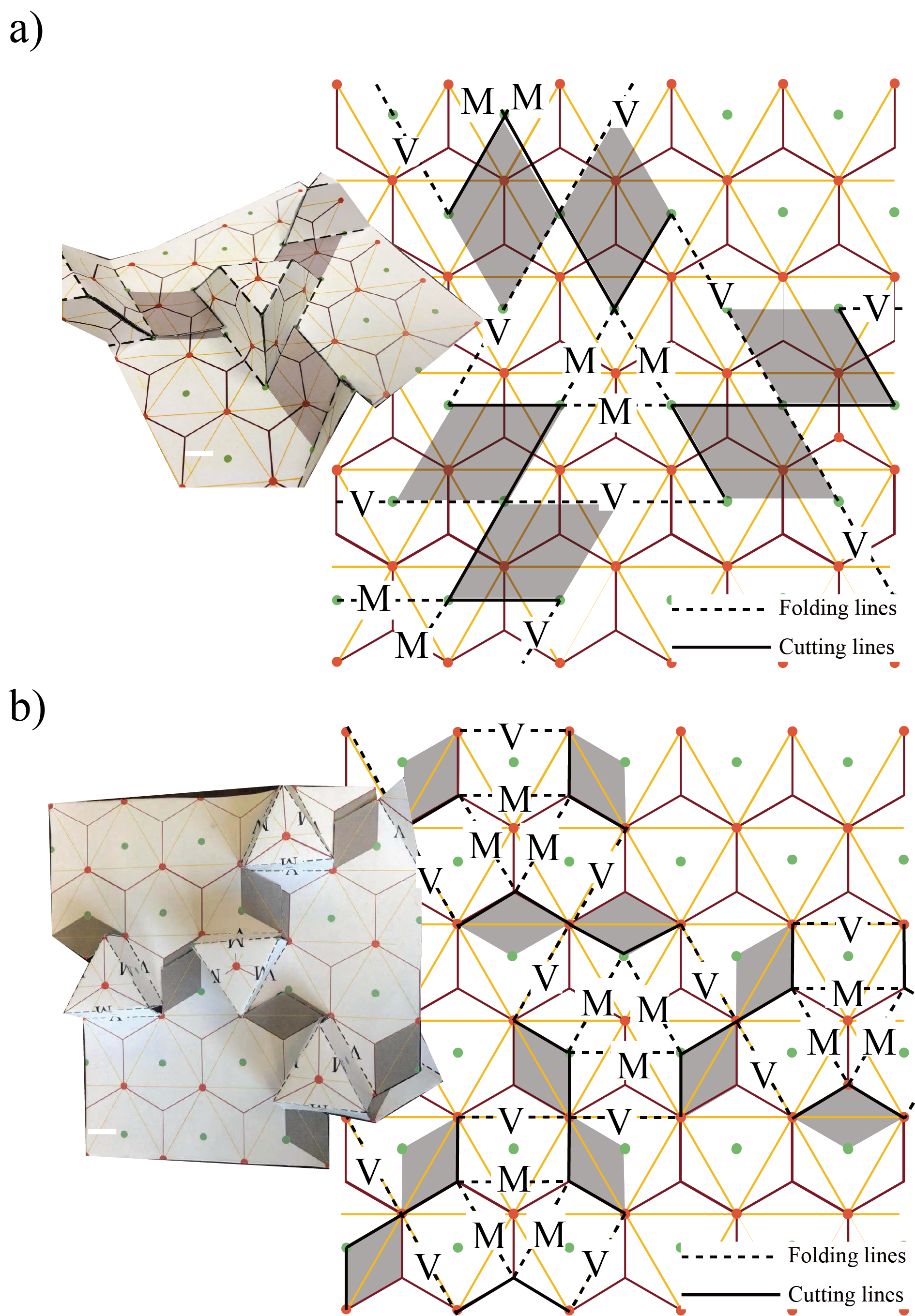}
	\caption{\label{fig:unitflat}
		A diagram of the interleaved {\sl kirigami} extension assembly showing the cooperative frustration between basic units. We cut along solid lines and make mountain (M) and valley (V) folds along dashed lines and, again,  the hatched areas overlap. a) Template for the three-fold structure with inset showing assembled pattern. ($\alpha=\pi/3$) b) Template for the three-fold structure with inset showing assembled pattern. ($\alpha=2\pi/3$) (scale bar: 1cm) 
	}
\end{figure}

We start with the honeycomb lattice and its dual lattice, the triangular lattice.  As in lattice {\sl kirigami} \cite{paper1}, which follows the insights so elegantly revealed in \cite{sadoc}, we create a dislocation in the lattice by introducing a disclination pair. { The parallel mountain and valley folds that create topography can be oriented at any angle with respect to the cuts. As shown in Fig.\ \ref{fig:basicmotif}, we label the interior angle of the polygon $\alpha$ and the angle between the cut and the fold $\beta$. Cutting $\alpha=\pi/3$ from one hexagon and combining two hexagons, each with $5\pi/6$ removed, creates a pentagon-heptagaon pair, a $\tilde{5}$-$\tilde{7}$ dipole (here and throughout, the tilde refers to the fact that these are defects on the dual lattice). The vector \textit{\textbf{l}} points from the dislocation to its anti-dislocation in each pair and \textit{\textbf{b}} is the Burgers vector of the dislocation.} {In contrast with previous lattice \textsl{kirigami} motifs, where \textit{\textbf{l}} and \textit{\textbf{b}} are parallel or perpendicular, here the two vectors are at $\pi/3$ with respect to each other.  Additionally, no material is removed along Burgers vector or the disinclination dipole.} When creating a plateau, as in Fig.\ \ref{fig:basicmotif}, we restrict ourselves for the moment, to regular, convex polygons. The discrete version of the Gauss-Bonnet theorem ({\sl i.e.}, geometry) requires that the cone angles add to $5\pi/3$ (here we have $\alpha+2\beta=5\pi/3$). For a single plateau, this does not constrain us. However, since we eventually consider periodic arrays, we require that the polygons respect the underlying lattice symmetry to preserve the intrinsic geometry of the lattice and dual lattice. It follows that $\alpha=\pi/3$ or $2\pi/3$ and so $\beta=2\pi/3$ or $\pi/2$, respectively.
 {The latter gives us the ``classic'' vertical walls in Fig.\ \ref{fig:unitflat}b and the former corresponds to new, tilted walls, with angle  $\theta=\sin^{-1}(1/3)\approx 19.47^\circ$ from the vertical.} 
 In general, for an isolated ``$\tilde 5$-butte'' with an internal angle $\alpha$, the sidewalls make an angle $\theta=\sin^{-1}[\tan(\pi/3-\alpha/2)\tan(\alpha/2)]$ with the vertical and so, in principle, we can consider buttes with interior angle up to $5\pi/6$ with inward tilting sidewalls. If we wanted angles larger than $\alpha=\pi$ we can accommodate them by inverting the structure -- that is, by putting a $\tilde 7$ defect on the top and a $\tilde 5$ on the bottom of the sidewall corners.

\begin{figure}
	\includegraphics[width=\linewidth]{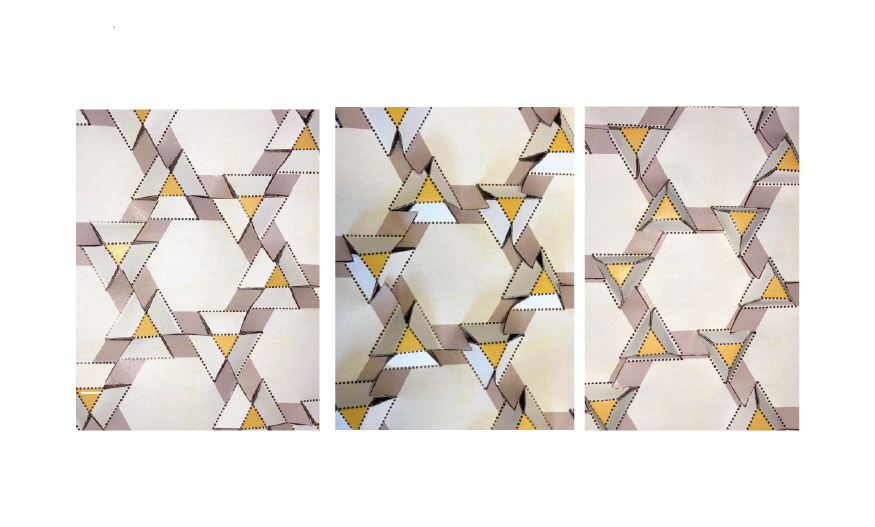}
	\caption{\label{fig:unitassemb}
From Left to Right: step by step assembly of the periodic plateau array.  Note that this assembly requires bending of the sheets and so this structure is {\sl not} rigid-foldable.  For clarity, the detailed pattern on  the underlying lattice is not shown.  Yellow areas are the plateaus/buttes and the gray areas overlap. (scale bar: 1cm)}
\end{figure}

In classic  {\sl kirigami}, paper can be slit and pulled, leading to out of plane distortion \cite{campbell, mc1, bert2} or, alternatively, paper is removed and rejoined to maintain piecewise flat panels \cite{paper1}. Here, however, we make the cuts but do not remove any paper,  yet we maintain piecewise flat geometry.  To make this possible we allow overlap between different parts of the same flat paper {\sl after folding}.  These paper extensions, as shown in Fig.\ \ref{fig:basicmotif}, neatly fit into the valley fold of the adjoining plateaus.  Although a single glide (like Fig.\ \ref{fig:basicmotif}) can hold its shape under small stresses, it can be unfolded easily along its dislocation direction.  However, in a triangular lattice of plateaus, unfolding requires that each unit relax in three dislocation directions.  By surrounding each unit with three other units as in Fig. \ref{fig:unitflat} and \ref{fig:unitassemb}, we can weave together the extensions to create a locked configuration -- unfolding in one direction is frustrated by compression in the other two.  

The design motif of self-locking assembly patterns can be extended to other lattices. As long as the two excess regions along the cuts have the same area, the flaps can overlap in the folded state and push against the buttes.  Friction between the paper or plastic deformation of the folds are necessary to hold the structure in place during the initial loading before the in-plane stress engages but not necessary: we will also discuss how plastically deformed materials can hold their initial shape as well.   The compact assembly brings extra connection between the edge of flaps and sidewalls, which can largely strengthen in terms of mechanical stiffness. As we will demonstrate, this special mechanical linkage leads to collossal specific strengths.  In the following, we will focus on this special mechanism and refer to each tetrahedral frustum as ``the basic unit.''  We will also modify our motif to square lattices in the next section (Fig. \ref{fig:square}). In this family, some patterns can be perfectly locked (no gap between hatched areas and folding lines), like the motif in  Fig.\ \ref{fig:unitflat}a.

\section{Strength Testing}
\subsection{The Original Model}
We fabricated the basic unit motif using four kinds of paper with different thicknesses (0.14 mm, 0.20 mm, 0.28 mm and 0.38 mm) \cite{papers}. {Additionally, we deployed our pattern on both transparency films (0.10mm) and copper sheeting (0.15mm) \cite{copperandphotofilm}. We did bending and tensile tests to characterize the materials we used in  Fig.\ \ref{fig:materials}.  

We begin our discussion with the thickness 0.28mm paper.}  We start with a letter-size sheet (215.9 mm by 279.4 mm) then cut and score it with a Graphtec CE6000 Series Cutting Plotter. Folding was performed by hand along the score lines.

\begin{figure}[t]
	\includegraphics[width=0.8\linewidth]{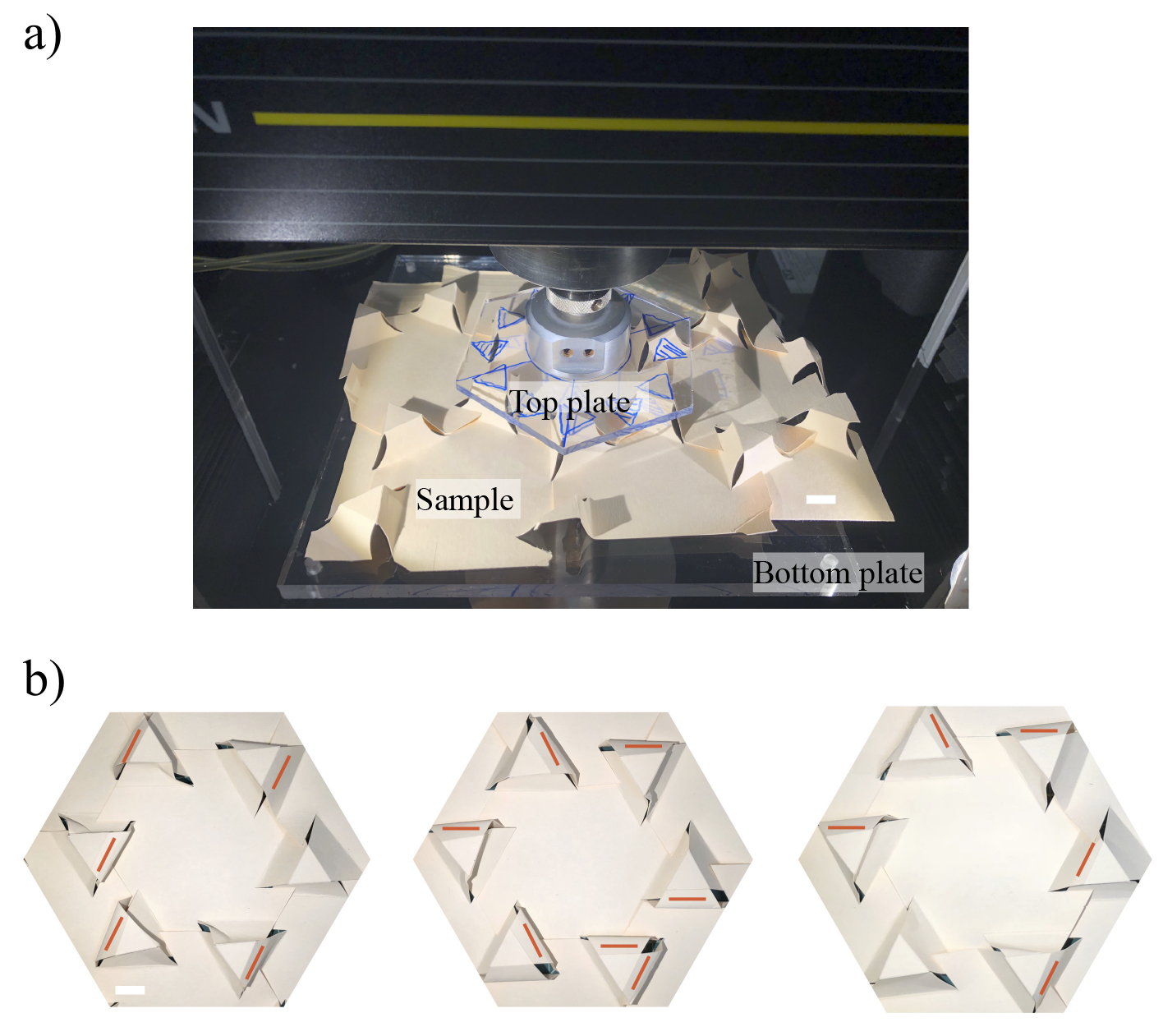}
	\caption{\label{fig:loading}
		The loading setup.  a) To avoid boundary effects, we only push on fully coordinated plateaus.  b) Upon collapse, some walls buckle in while others buckle out.  We denote the outward buckled walls with red lines.  Three different samples show that the collapse has a random component, likely due to folding inhomogeneity. (scale bar: 1cm)}
\end{figure}

To assess the strength of each assembly, we performed tests on the Instron 5564 with a 2kN load cell to get force-displacement curves.  Throughout, we set a loading rate of 2mm/min.  We sandwiched each sample between thick, acrylic plates (the top hexagon plate mass is 61.9g) to spread the stress evenly. The data collection was performed by Series X in Merlin software.  The loading setup is shown in Fig.\ \ref{fig:loading}a.

 \begin{figure}[h]
	\includegraphics[width=0.95\linewidth]{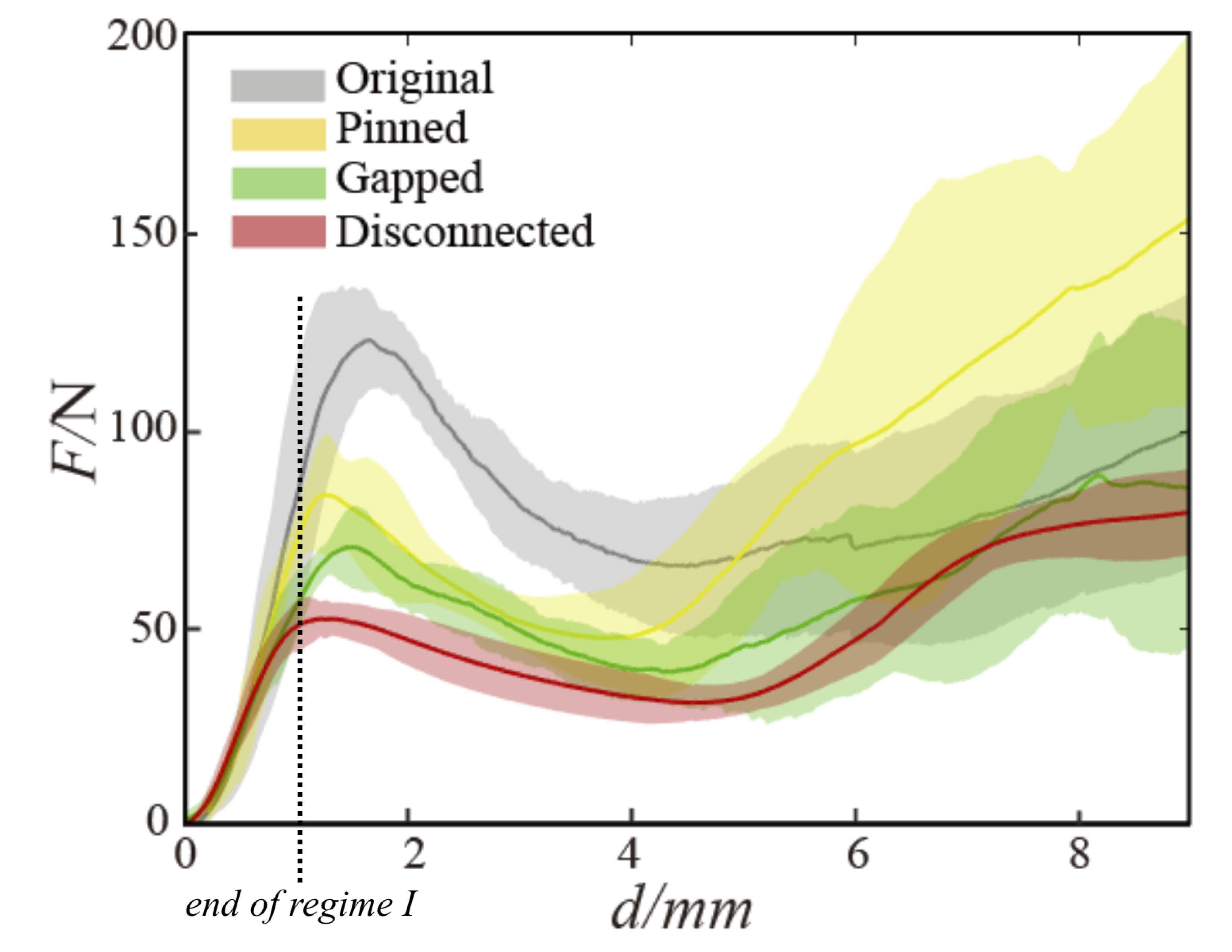}
	\caption{\label{fig:pinnedcompare}
		Experimental curves for the original, pinned, gapped, and disconnected models. Note that the pinned and gapped are nearly identical: in one case there is no in-plane compression and in the other, the adjacent edges fail to support each other.   We have delinieated the end of Regime i) with the vertical dashed line -- it is the region where all four models (pinned, gapped, original, and disconnected) have the same response.  At larger strain, the ``unoriginal'' models fail while the original model continues to resist.
	}
\end{figure}

The load-displacement curves are shown in  Fig.\ \ref{fig:pinnedcompare}a with the gray region representing the error bars from  {five} samples.  We observe that the loading curve has three regions: i) the initial linear regime where the load cell and six frustums are getting into full contact and each butte is independent; ii) a new (reproducible) regime where the {\it collective} structure is responding linearly; iii) the peak and beyond where the details of wall-buckling give huge variations in response -- note that the final rise in force comes from the crushing of the crumpled paper. The distinction between the first and second regions is seen clearly in Fig. \ref{fig:pinnedcompare}: there we see that isolated, pinned, and gapped buttes have the same response as the original model in regime i) and then collapse at some point.  Regime ii) is the``post-single-butte'' response and is only present in the original model.
{Error in the linear regions i) and ii) arises from the initial gauge length, which is not an intrinsic property of the structure and the load curves are reproducible up until the peak.  For each experimental set of parameters (geometry, pinning, paper thickness {\sl etc}), the data from different samples can be collapsed onto one curve by dividing \textsl{F} and \textsl{d} with $\textsl{F}_{peak}$ and $\textsl{d}_{peak}$, respectively.  The small variations of force peaks from sample to sample are mainly induced by the environment humidity and manual inaccuracy. For the former, paper stiffness and surface friction drop with the increase of water content in wooden fibers ~\cite{paperph1} and we observed seasonal variation in the peak load.  In order to ameliorate this effect we baked all models before loading.  Regime {ii)}, however, is sensitive to the roughness of the paper.  We have tried other materials such as smooth transparency sheets (that are stiffer than the original paper) and find that the failure of those structures is related to low friction: they returned back to their flat states once being pushed {\sl before} the in-plane stresses locked the plateaus.}  As shown in  Fig.\ \ref{fig:othermaterial}, our geometry can be transferred to other materials. We will discuss both paper and material differences later.  We also note that when we taped the overfolded flaps to the underlying material, we {\sl did not} see an increase in strength -- were friction the controlling parameter, presumably the ``infinite'' friction provided by the tape would lead to even stronger structures (see Fig. \ref{fig:pinned}).

\begin{figure}[t]
	\includegraphics[width=\linewidth]{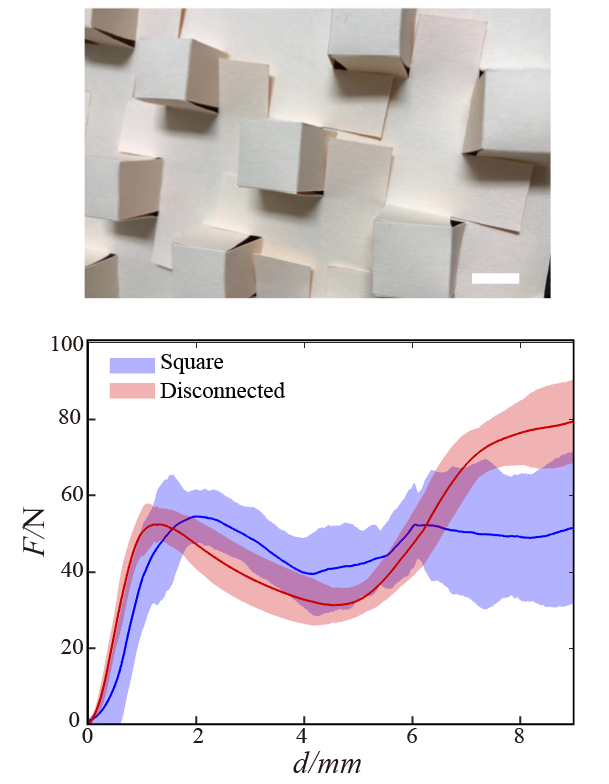}
	\caption{\label{fig:square}
		Square interleaved {\sl kirigami} edge assembly.  Top shows the interlocking pattern. On bottom, we see  that the vertical walls do not convert downward force to in-plane force, preventing the collective response of the design, similar to the disconnected model.  We pushed on four square sites with edge length 17.3mm compared to six equilateral triangle sites with the same edge length so the square model has an even larger contact area by a factor of $8\sqrt{3}/9\approx 1.6$. (scale bar: 1cm)
	}
\end{figure}

\begin{figure}[b]
 	\includegraphics[width=0.9\linewidth]{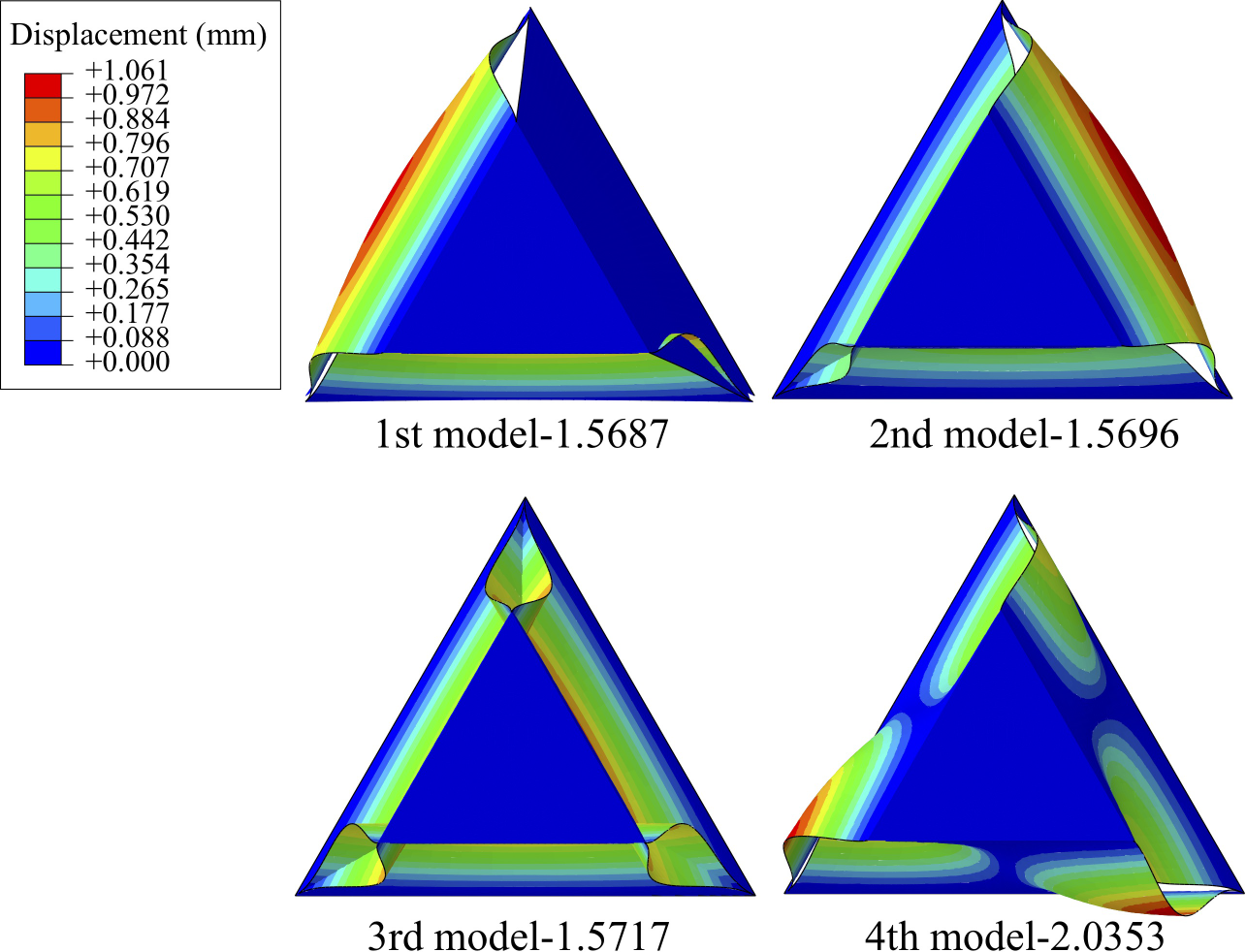}
 	\caption{\label{fig:fem}
 		Numerical images of the first four eigenmodes and the corresponding eigenvalues. The deformation level in images is enlarged by a factor of four for clarity. 
 	}
 \end{figure}

Finally, note that the  variation from sample to sample becomes large after  buckling (Fig. \ref{fig:loading}b)). We attribute this to hard-to-control variations in folding and cutting leading to ``random'' buckling of the sidewalls, either in or out.  This sensitivity arises from the nearly degenerate deformation modes of the structure: we used  FEM to calculate the linear buckling modes of thin plates organized into the original model.  As shown in Fig. \ref{fig:fem}, the inward and outward buckling modes have nearly degenerate eigenvalues and this is why small variations (imperceptible to us) change the nature of buckling (in or out). The first three modes were all bending modes with very close eigenvalues (Fig.\ \ref{fig:fem}).  The fourth mode was a twisting mode with a larger eigenvalue gap.  Note also, that in the FEM analysis, different buckling in the individual sidewalls does not seem to have an effect on neighboring walls -- they buckle independently.  Together, these simulation results  rationalize the unpredictable variation in the post-peak regime.  
 
 \begin{figure}[t]
 	\includegraphics[width=0.9\linewidth]{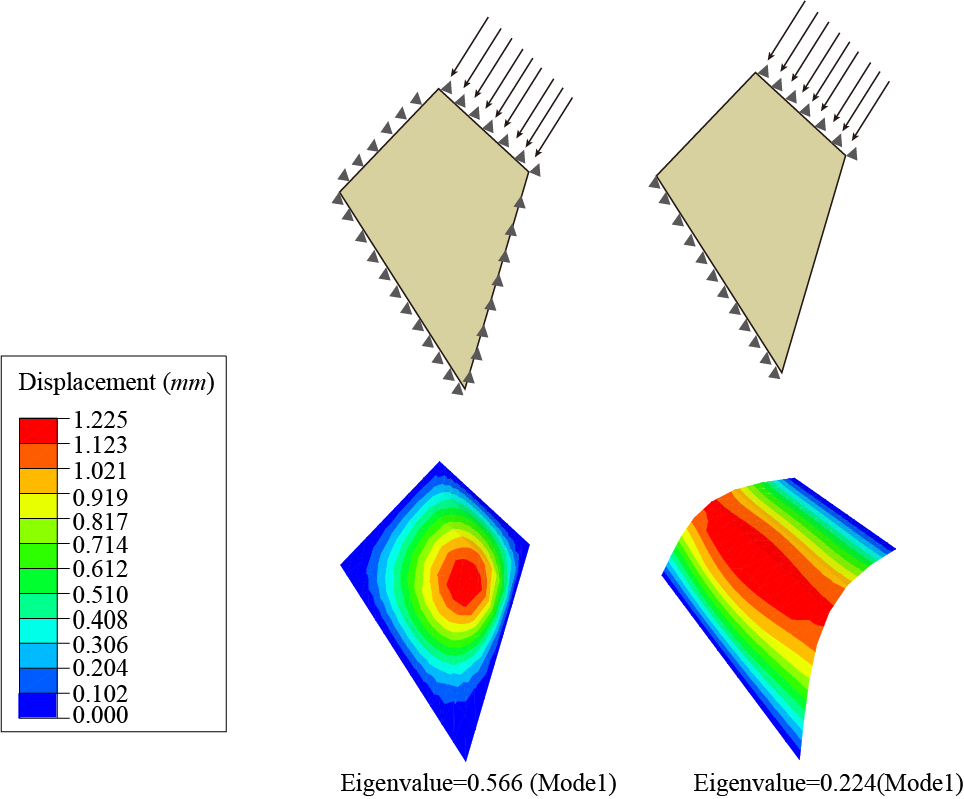}
 	\caption{\label{fig:walls}
 		Numerical buckling images of a sidewall with and without support at two isosceles edges. Solid triangles around the boundary are translational constraints.  The sheets are completely restrained along the three longest edges while the shortest edge is allowed to move along the loading direction as depicted by the arrows. The deformation level in images is enlarged by a factor of four for the reader's pleasure. 
 	}
 \end{figure}
 
 \subsection{Probing the Source of Strength through Structural Modifications}
\begin{figure}[h]
	\includegraphics[width=0.9\linewidth]{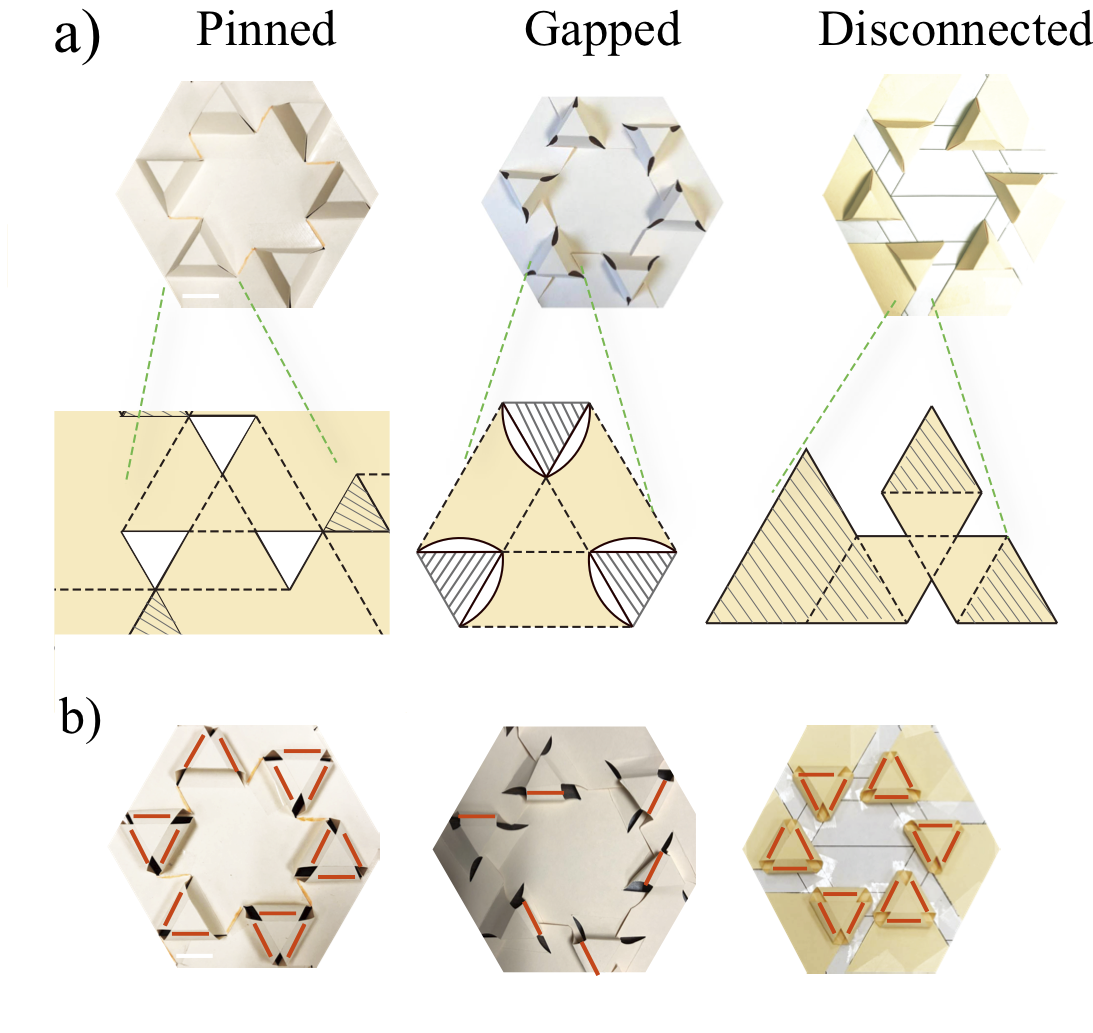}
	\caption{\label{fig:pinned}
		Pinned, gapped, and disconnected models.  a) Diagrams of cuts and folds. For pinned models, we made cuts along solid lines and added tape on the hatched areas to hold the shape. For the gapped models, we removed the edges of the sidewalls to prevent them from touching adjacent walls upon compression.  For disconnected models, we folded six units separately and stuck them at the positions they would sit at were they on the original model. b) Final states of all three models. Red short lines denote walls buckled outside.  Note that outward buckling was the dominant mode in the pinned and disconnected cases. (scale bar: 1cm)	}
\end{figure}

 To fully explore how neighbors contribute to the overall strength of the structure, we compared the original models (just described) with both pinned models and disconnected models. In the pinned models, we removed the influence of the excess flaps by pinning the flaps with tape as on the left in Fig. \ref{fig:pinned}.  The tape does not modify the units, each trapezoidal edge remains free; the tape only keeps the flaps from moving and prevents the conversion of downward forces to in-plane compression.  Likewise, on the right of Fig. \ref{fig:pinned}  we physically disconnected the plateaus to remove all interactions.  All the models here were made with the 0.28mm paper.

\begin{figure}[t]
	\includegraphics[width=0.9\linewidth]{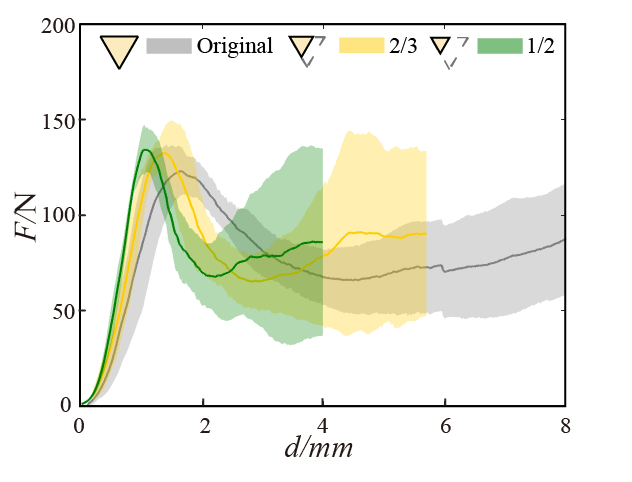}
	\caption{\label{fig:density}
		The influence of structure density. The models were modifed by changing the dimensions of the basic units as depicted and we only pushed on six, fully-coordinated units.  Note that the yield force was independent of the dimension implying that the yield pressure scaled as the inverse of the square of the length.  (The top plates were all different mass: 69.1g, 25.3g, and 9g, for the original, 2/3, and 1/2 scale structures, but this difference is well within the noise of the measurement).   
	}
\end{figure}

As expected, the disconnected model is the weakest.  However, as we alluded to in the last section, the original model is {\sl stronger} than the pinned model by roughly a factor of two.  What accounts for this?  We note that the transfer of vertical stresses on the plateaus into in-plane stresses on the structure can only occur when the flaps are both present and are free to move.  Both are needed for a structure to have a ``Regime ii)'' where collective effects ensue. This creation of compression locks together the plateaus along their open edges transforming them from three separate Euler elastic beams into a rigid frustum-shaped polyhedra with freely-supported faces.  The strength of the original structure is created by the mutual support of one wall by its two adjacent partners. The roughness of and friction between the slits confers more or less strength to the design in the initial loading.   However, as we will discuss further in \S{\bf IV} , both friction between the overlapping regions and plastic deformations of the folding hold the structure in place -- the less the structure slips before the in-plane compression starts to grow, the more resilient the structure.    Indeed, in terms of fabrication, smooth paper is inferior to rougher paper. Further, note that in the pinned and disconnected models the dominant buckling mode is outwards but the freely-supported plate model requires {\sl inward} buckling. Fortuitously, the inward buckling is promoted by the displacement of the flaps.  Before moving on we also note that the data in Fig. \ref{fig:pinnedcompare} suggests that collective randomness in the buckling modes is at least partly responsible for the large variation of the load response for large displacement: there is much less scatter in the disconnected models than in any of the models where the plateaus interact.

To further test the in-plane coupling hypothesis, we constructed a square analog of the original model. As shown in Fig. \ref{fig:square}, this model has square, overlapping flaps but has vertical sidewalls.  Without any in-plane stresses, vertical sidewalls should be stronger than canted sidewalls since the sloped walls of the original model must bear larger compression for the same load.  However, we do not find this.  The square model is weaker, presumably because there is no transfer of load to in-plane compression and thus the square plateaus do not get compressed and behave as a solid, cubical shell.

\subsection{Modeling Simple Supports}

{From the point of view of each individual wall, the inward deformation transforms a free-standing Euler beam into a plate with four simply supported edges -- edges that are constrained to have zero displacement but for which bending is allowed.  Given the continuity of plates around each butte, each wall can carry a much higher load than an isolated Euler strip~\cite{stable}.  The finite element method (FEM) results from general implicit code ABAQUS/Standard verify that the subsequent interaction between walls leads to higher strength. As shown in Fig. \ref{fig:walls}, we tuned the boundary conditions at the two tilted unconnected edges  from ``free edges'' (no constraint) to ``simply supported edges'' (only allowing rotation) to simulate the contact behavior. The compression tests were simulated by imposing load at the top edge. This modification leads to a factor of 2.5 higher buckling threshold.}

In experiment, we removed the interior reinforcement inside each basic unit and created ``gapped'' models where edge contact was avoided by cutting some of the paper along the edges of the sidewalls (also shown in Fig. \ref{fig:pinned}). The gapped model showed nearly the same loading curve as the pinned model. In the former case, the sidewalls are still Euler beams that do not contact and form shells, in the latter, there is no in-plane compression to press the walls together.  Therefore both of these loading curves are giving us information about the Euler-like buckling of the walls which are, appearently,  {\sl identical}.  
{ From Fig. \ref{fig:loading}b), just around 2/3 walls buckle inside and form contact shell. Correspondingly, the buckling peaks of the original mode are about 2 times higher than that of gapped models in Fig. \ref{fig:pinnedcompare}, not far from our FEM estimate of 2.5. 
 Also, comparing the disconnected models with the gapped models, we see a stiffness increase originating from a ``neighbor effect" with a factor of 1.4. }   	
\begin{figure}
	\includegraphics[width=0.9\linewidth]{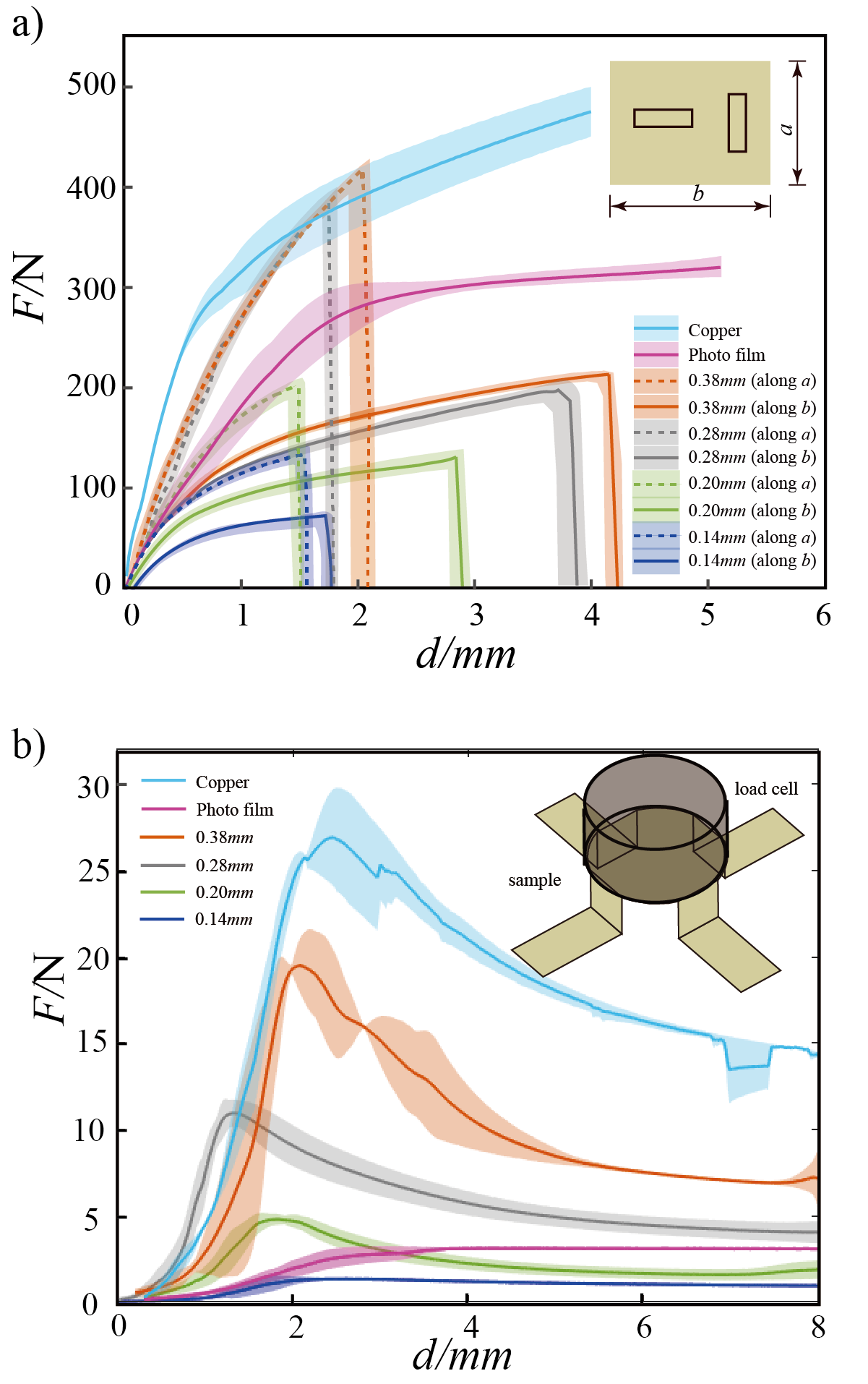}
	\caption{\label{fig:materials}
		The characterization of materials. a) Tension tests of 30mm by 100mm strips. We cut strips along orthogonal directions. Paper is the only material showing anisotropic behavior. We note that copper, in particular, undergoes plastic deformation under tension as discussed in \cite{plasticpaper}. b) We captured the bending behavior of soft and inhomogeneous materials by using four-legged frames (insert: experimental setup). The legs were all 16mm by 16mm squares.  }
\end{figure}

\begin{figure}[h]
	\includegraphics[width=0.9\linewidth]{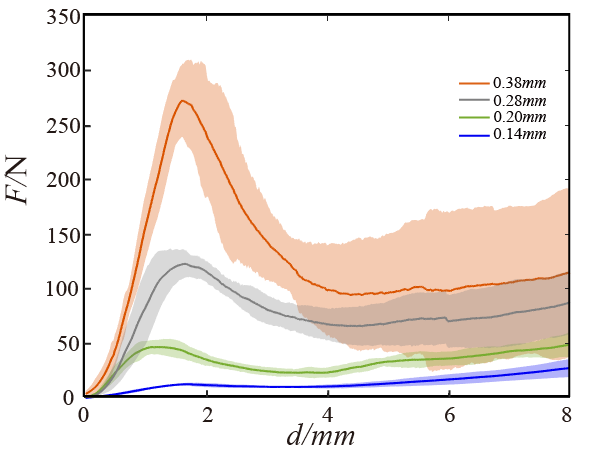}
	\caption{\label{fig:thickness}
		The influence of paper thickness. Experimental curves of paper with  thickness  0.14mm, 0.20mm, 0.28mm (original models) and 0.38mm. 
	}
\end{figure}

\begin{figure}[h]
	\includegraphics[width=0.9\linewidth]{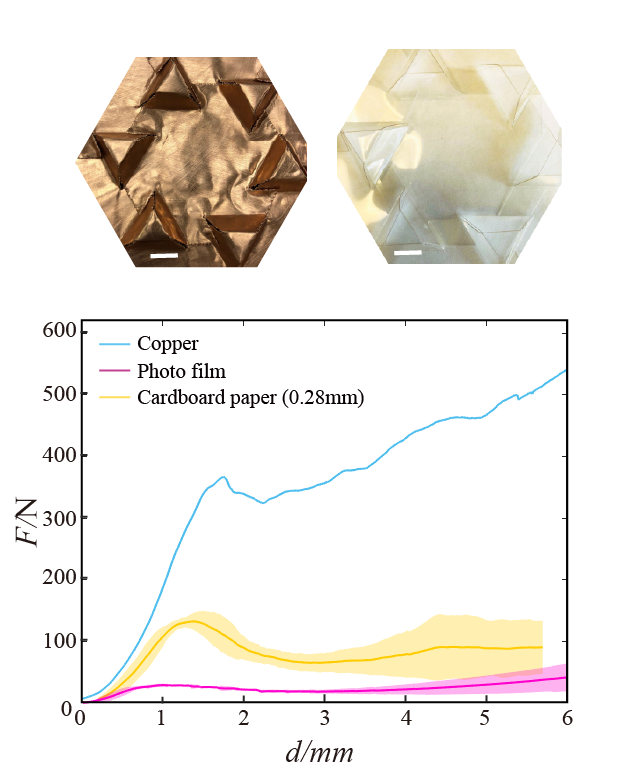}
	\caption{\label{fig:othermaterial}
		The influence of material difference (2/3 scale original structure for copper, photo film, and paper). Top: copper model (scratched) and polymer film model (unscratched). Bottom: experimental curves of copper sheet, transparency films and paper with thickness 0.28mm. (scale bar: 1cm) 
	}
\end{figure}

We made measurements with different unit sizes, scaled by the linear factor $\lambda$.  We always pushed on six, fully-coordinating units.  Since the contact area shrank as $\lambda^{2}$,  but, as Fig. \ref{fig:density} shows,  the buckling force remained {\sl constant} , the buckling pressure scales as $\lambda^{-2}$.  As we shall see in \S{IV}, this is consistent with the theory of thin plates and suggests that the strength is coming from the length of the sidewalls and not the spacing between buttes (though the latter does contribute to the load per unit area).  In closing this section, we note that a full sheet of 1/2 scale units has a mass of 8.1g and buckles at around 1100 N, a load-to-weight ratio around 14000!

\section{Other Materials and the Role of Friction and Plasticity}

Finally, we address the material dependence of the behavior of the assemblies. {To better understand the material properties, we performed tensile and compression tests on copper sheets, transparency films, and four kinds of paper.} We note that paper is a complicated material: as shown in Fig.\ \ref{fig:materials}a), it is not isotropic -- there is a machine rolling direction and the transverse direction~\cite{paperph2}. Fibers are primarily aligned along the machine direction, leading to a mechanical difference in different directions. However, the structures we consider have three fold symmetry, washing out the anisotropy of the paper -- we checked that the buckling thresholds are independent of orientation (within experimental error).  Somewhat surprisingly, from Fig.\ \ref{fig:materials}a), we observe that the 0.28mm and 0.38mm paper showed almost the same loading curves. This is consistent with the construction of cardboard: it is known that the stiffness of cardboard (0.28 mm and 0.38 mm paper are both cardboard) is controlled by the top and bottom surface layers, made of chemical pulp while the mechanical pulp -- the filler that makes up the rest of the material -- is sandwiched between~\cite{paperph2}. We also verify that copper and transparency photo films are ductile under tension while paper and cardboard are brittle.

{We employed a four-legged symmetric setup in Fig.\ \ref{fig:materials}b) to average between the strong and weak direction of anisotropic paper.} The 0.28mm and 0.38mm paper have the same Young's modulus \textsl{E} from tensile tests, implying that the paper is not an isotropic material. Note that the bend stiffness of thin plates (with two free edges) is $\textsl{D}=\textsl{E}\textsl{h}^{3}/[12(1-\nu ^{2})]$ where \textsl{h} is paper thickness and $\nu$ is Poisson ratio. Moreover, the critical load per unit length is $N_{cr}=\pi^{2}D/l^{2}$ where \textsl{l} is the height of plates \cite{stable}. Therefore, with the same $E$, the ratio of peak strengths of the 0.38mm and 0.28mm paper should be $(38/28)^{3}\approx2.4$, which can be verified in Fig.\ \ref{fig:materials}b) and Fig.\ \ref{fig:thickness} where we find a ratio around 2.    The consistency of the theory with the measurement corroborates the hypothesis that paper is a complex material.  Finally,   {As shown in Fig.\ \ref{fig:thickness}, the 0.14 mm paper is much softer than the other three.}  Because it is thin its behavior is in the membrane regime where compression is difficult to support, {\sl i.e.} $a/h\approx 15/0.14 > 80$ where $a$ is the dimension along which we bend~\cite{book}. 

{ Could we, indeed, fold metal sheet into this motif? }  We tried both aluminum and copper foil.  Our construction is not rigid-foldable and requires some bending of the flaps to be built.  We found that aluminum foil was too brittle, but copper, being quite ductile could be manipulated into our motif.  Our cutting machine was unable to cut through the 0.15mm copper roll \cite{copperandphotofilm} but, using push pins, we were able to cut the lines manually and then fold our motif as shown in Fig.~\ref{fig:othermaterial}.  Unlike paper, even smooth copper can hold its folded shape easily due to the ease of plastic deformation.  
However, to facilitate assembly, we also roughened the copper surface with 120 grit sandpaper and we recapitulate our original observations on the cardboard -- three
 regimes, with the second one coming from the collective strength of the neighbor's effect, in-plane compression and simply-supported Euler plates.   
 
 Does our data in Fig. \ref{fig:othermaterial} demonstrate the both copper and photo films, folded into the original motif, assume the same ``super-strength'' displayed by the paper?  Using the measurements of the bending modulus in Fig. \ref{fig:materials}b) we can estimate the ratio of the bend moduli of photofilm to the 0.28mm paper as the ratio of the critical loads ($N_{cr}=\pi^{2}D/l^{2}$), roughly 0.25, in line with the estimate from Fig. \ref{fig:othermaterial} of about $1/5$.  We can do the same estimate with the copper data from Fig. \ref{fig:materials}b) and would expect an increase in strength of roughly 3 if the copper model has lock-in.   Indeed, Fig. \ref{fig:othermaterial} demonstrates a comparable ratio.  While the agreement is suggestive, we note that copper foil is highly plastic \cite{plasticpaper}, and elasticity measurements are delicate.  In fact, our measurement of the Young's modulus in Fig. \ref{fig:othermaterial}a) is off by an order of magnitude from what one would expect from the reported value of $E\sim 120GPa$. 

We also tried cutting and scoring transparency films. We used totally scratched, partially scratched (only walls), and unscathed photo films  \cite{copperandphotofilm} to make the motif. The three showed no difference in force-displacement curves. Models made by original smooth films can reach the same force peak if, upon folding, we create plastic deformations that allow the films to retain their shape with zero load (without doing this, we found that the bending modulus of the transparencies was enough to unfold them).  Although structural geometry, rather than material friction, conspires to create colossal strength, we can save much manual effort by folding rough materials and by keeping the structure folded throughout deformation, friction can guarantee that the force peak is attained.  What this demonstrates is that there must be some way of holding the structure together up until the in-plane compression generates the freely-supported plateau plates.

\section{Summary and Conclusions}
In summary, we have designed an interleaved {\sl kirigami} extension assembly and characterized its mechanical properties. Not only does our pattern  hold its shape through overlapping frictional flaps or plastic deformation of the material, but it conducts forces effectively between adjacent units through in-plane compression.  We have suggested that, in turn, the compression forces dramatically strengthen the structures .A variety of tests confirm this hypothesis.  It would be interesting to consider other overfolded geometries to create materials that could withstand large shears or bending out of plane: in this case  the role of the extensions becomes more subtle as they must respond and distort in a way that maintains the integrity of the overlapping motif.  We leave this for future work.

\begin{acknowledgments}
We thank H. Ansell, J. Li, O. Parrikar, A. Seddon, and M. Tanis for discussions and the insights of the referees. We thank Y. Gao for assistance with the laser cutter and the Instrom.  This work was supported by NSF DMR12-62047 and a Simons Investigator grant from the Simons Foundation to R.D.K.   X.W. was also supported by the China Scholarship Council.  R.D.K. would like to thank the Isaac Newton Institute for Mathematical Sciences for support and hospitality during the programme The Mathematical Design of New Materials when work on this paper was undertaken.  This work was also supported by EPSRC Grant Number EP/R014604/1.  
\end{acknowledgments}

\clearpage


\begin{thebibliography}{10}
\bibitem{papers} In particular, the 0.14 mm is 24 lb Mohawk Via Laid, Bright Wight paper, the 0.2 0mm is Strathmore (brand) 400-4 series acid free feuilles drawing paper (80 lb), the 0.2 8mm is cut from Office Depot (brand) 1/3 cut tab 30\% recycled manila file folder, and the 0.38 mm is Canson(brand) XL recycled Bristol pad (95 lb).
\bibitem{copperandphotofilm} In particular, the photo film is 18 lb 3M (brand) transparency film for Canon $\&$ Epson Ink Jet Printers , the copper roll is made of 99.9$\%$ pure copper alloy and maufactured to specifications of ASTMB-152 from Basic Copper(brand).
\end{thebibliography}
\end{document}